\documentclass[journal]{IEEEtran}

\usepackage{amsmath}    
\usepackage[pdftex]{graphicx}                %图片
\usepackage{amssymb}                         %输入实数集，复数集这样的符号
\usepackage{booktabs}                        %画三线表
\usepackage{bm}                              %专门处理数学粗体的bm宏包
\usepackage{color}
\usepackage{enumerate}                       %用于enumerate结构
\usepackage{epstopdf}                        %用于使用eps图片
\usepackage{mathtools}                       %排版公式字符的间距
\usepackage[ruled,norelsize]{algorithm2e}    %公式，见https://tex.stackexchange.com/questions/147598/how-to-use-the-algorithm2e-package-with-ieeetran-class
\ifCLASSOPTIONcompsoc            %画subfigure
  \usepackage[caption=false,font=normalsize,labelfont=sf,textfont=sf]{subfig}
\else
  \usepackage[caption=false,font=footnotesize]{subfig}
\fi

\makeatletter
\newcommand{\removelatexerror}{\let\@latex@error\@gobble}
\makeatother

\begin{document}

\title{Towards Ubiquitous Positioning by Leveraging Reconfigurable Intelligent Surface}

\author{Haobo~Zhang,~\IEEEmembership{Student Member,~IEEE,}
        Hongliang~Zhang,~\IEEEmembership{Member,~IEEE,}
        Boya~Di,~\IEEEmembership{Member,~IEEE,}\\
        Kaigui~Bian,~\IEEEmembership{Senior Member,~IEEE,}
        Zhu~Han,~\IEEEmembership{Fellow,~IEEE,}
        and~Lingyang~Song,~\IEEEmembership{Fellow,~IEEE}% <-this % stops a space
\thanks{H. Zhang and L. Song are with Department of Electronics, Peking University, Beijing, China (e-mail: \{haobo.zhang,lingyang.song\}@pku.edu.cn).}%
\thanks{H. Zhang is with Department of Electronics, Peking University, Beijing, China, and also with Department of Electrical Engineering, Princeton University, NJ, USA (e-mail: hongliang.zhang92@gmail.com).}%
\thanks{B. Di is with Department of Electronics, Peking University, Beijing, China, and also with Department of Computing, Imperial College London, London, UK (e-mail: diboya92@gmail.com).}%
\thanks{K. Bian is with Department of Computer Science, Peking University, Beijing, China (e-mail: bkg@pku.edu.cn).}% 
\thanks{Z. Han is with the Department of Electrical and Computer Engineering, University of Houston, Houston, TX 77004, USA, and also with the Department of Computer Science and Engineering, Kyung Hee University, Seoul 17104, South Korea (e-mail: zhan2@uh.edu).}% <-this % stops a space
}

\maketitle

\begin{abstract}
  The received signal strength~(RSS) based technique is widely utilized for ubiquitous positioning due to its advantage of simple implementability. However, its accuracy is limited because the RSS values of adjacent locations can be very difficult to distinguish. Against this background, we propose the novel RSS-based positioning scheme enabled by reconfigurable intelligent surface~(RIS). By modifying the reflection coefficient of the RIS, the propagation channels are programmed in such a way that the differences between the RSS values of adjacent locations can be enlarged to improve the positioning accuracy. New challenge lies in the selection of suitable reflection coefficients for high-accuracy positioning. To tackle this challenge, we formulate the RIS-aided positioning problem and design an iterative algorithm to solve the problem. The effectiveness of the proposed positioning scheme is validated through simulations.
\end{abstract}

\vspace{-2mm}
\begin{IEEEkeywords}
  Ubiquitous positioning, reconfigurable intelligent surface, received signal strength.
\end{IEEEkeywords}

  % \ifCLASSOptIONpeerreview
  % \fi
\IEEEpeerreviewmaketitle

\vspace{-3mm}
%%%%%%%%%%%%%%%%%%%%%%%%%%%%%%%%%%%%%%%%%%%%%%%%%%
\section{Introduction}
%%%%%%%%%%%%%%%%%%%%%%%%%%%%%%%%%%%%%%%%%%%%%%%%%%
\label{introduction}
\vspace{-1mm}

The increasing demand for location-based applications such as navigation, healthcare monitoring, and indoor positioning has led to a growing interest in ubiquitous positioning or positioning anywhere. Among various kinds of positioning techniques, received signal strength~(RSS) based technique is widely used because it can be easily implemented on the widespread Wi-Fi compatible devices with little hardware requirements~\cite{yassin2017recent}.

In the literature, various RSS based techniques have been discussed. For example, the authors in~\cite{bahl2000radar} proposed a deterministic location estimation method using the RSS measurements. The probabilistic methods were adopted in~\cite{youssef2008the} to infer the user's location. In~\cite{ouyang2010received}, the authors studied the positioning of a group of wireless sensor nodes, and the convex location estimators were utilized. However, the positioning performances in the aforementioned works highly depend on the RSS distribution which can degrade the positioning accuracy if unfavorable. Specifically, in unfavorable RSS distributions, the RSS values at different sampling locations are close, and thus, these locations are difficult to be distinguished.

Recently, the reconfigurable intelligent surface~(RIS) has been proposed as a promising solution to actively customize the radio environment~\cite{di2019hybrid}. An RIS is a planar surface consisting of many reflecting elements, which can be coated on the surface of various objects such as walls~\cite{Mohamed2020Recongurable}. The reflection coefficient of the RIS can be adjusted by changing the elements' states, which are defined as the RIS configuration~\cite{zhang2020reconfigurable}. Thus, we can customize the RSS distributions in the radio environment by setting RIS configurations, which implies
that the positioning accuracy can be potentially improved by integrating RIS into the positioning scheme.

In this letter, we consider the RSS based multi-user positioning enabled by the RIS. An access point~(AP) emits signals that are reflected by the RIS, and users measure the RSS values for localization. Different from traditional RSS based techniques, the RSS distribution can be adjusted by changing the RIS configuration. Therefore, we can select suitable configurations to enlarge the RSS differences at different locations for high-accuracy positioning.

However, it is challenging to select suitable configurations due to the enormous number of configurations and the complicated relation between the configuration and the RSS distribution. To tackle this challenge, we formulate the positioning problem to minimize the weighted probabilities of false positioning, and design the configuration optimization~(CO) algorithm to efficiently solve the formulated problem.

\vspace{-4mm}
%%%%%%%%%%%%%%%%%%%%%%%%%%%%%%%%%%%%%%%%%%%%%%%%%%
\section{System Model}
%%%%%%%%%%%%%%%%%%%%%%%%%%%%%%%%%%%%%%%%%%%%%%%%%%
\label{s_system}
\vspace{-2mm}

%\vspace{-1mm}
\subsection{Ubiquitous Positioning Scenario}
\vspace{-2mm}

\begin{figure}[!t]
  \setlength{\abovecaptionskip}{-3pt}
  \setlength{\belowcaptionskip}{-25pt}
  \centering
  \includegraphics[width=2.6in]{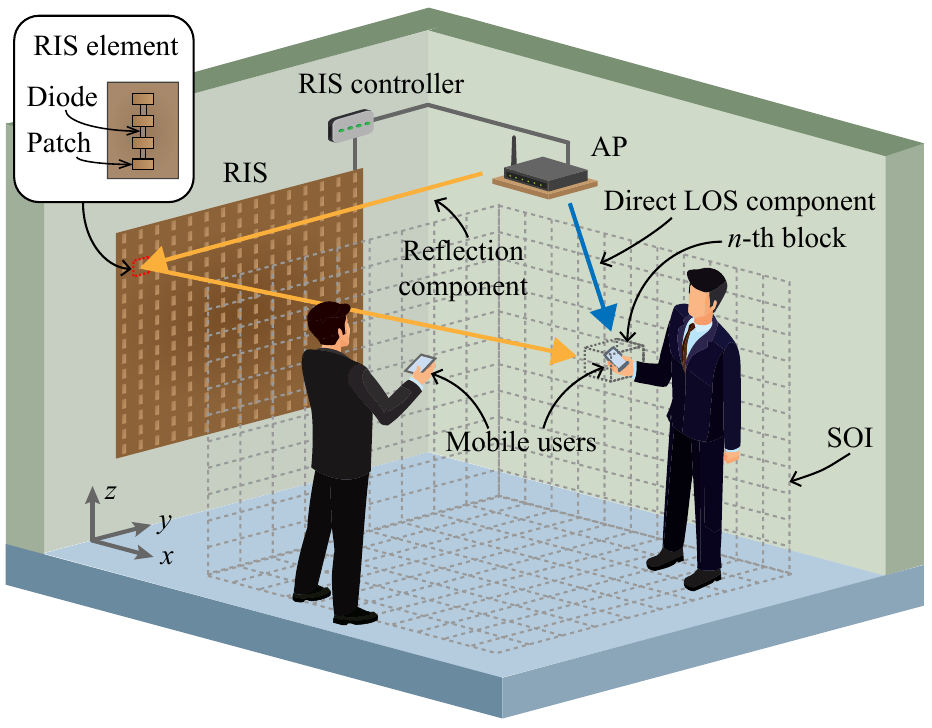}
  \caption{System model for the RIS aided multi-user positioning.}
  \label{f_system}
  \vspace{-3mm}
\end{figure}

As shown in Fig.~\ref{f_system}, we consider an indoor positioning scenario consisting of an AP, an RIS, and multiple users requiring their own location information. The AP connects to the RIS controller which regulates the operation of the RIS. During the positioning process, the AP sends single-tone signal over frequency $f_c$ to the RIS and users, and the RIS reflects the signals from the AP to the users. Each user measures the RSS for positioning. This positioning scheme can also be implemented in other scenarios by coating the RIS on the wall and placing an AP nearby.

To be specific, we assume that all users move in a cubic region of size $l_x\times l_y\times l_z$, which is referred to as the space of interest~(SOI). The SOI is discretized into $N$ blocks with the same size denoted by $\mathcal{N} = \{1, \cdots, N\}$. The user's location can be represented by the index of the block where it is. Since the RIS is able to customize the reflected signals, it can be used to control the RSS at different blocks and improve the positioning accuracy of the RSS based technique.

\vspace{-4mm}
\subsection{RIS Model}
\label{RIS}
\vspace{-1mm}

The RIS is an artificial material which is composed of a two-dimensional array of metal and dielectric elements, as illustrated in Fig.~\ref{f_system}. In each element, there are several subwavelength-scale metal patches connected by PIN diodes on the dielectric substrate. By adjusting the bias voltages of the PIN diodes, the reflection coefficient of the element can be changed. Here, the reflection coefficient is defined as the ratio of the reflected signals to the incident signals.

The RIS consists of $M$ elements which are denoted by $\mathcal{M} = \{1, \cdots, M\}$. Each element has $C$ reflection coefficients with uniform phase shift interval $\Delta\theta = 2\pi/C$ and phase-dependent amplitudes~\cite{abeywickrama2020intelligent}. Thus, the reflection coefficient of element $m$ can be expressed as
\vspace{-2mm}
\begin{equation}
  r_m(c_m) = r(c_m) e^{-jc_m\Delta\theta},
  \vspace{-2mm}
\end{equation}
where $r(c_m) \in [0, 1]$, and $c_m \in \{1, \cdots, C\}$. For convenience, we refer to $c_m$ as the state of the $m$-th element. Besides, the \emph{configuration} is defined as the vector of all the RIS elements' states which is denoted by $\bm{c} = (c_1, \cdots, c_M)$.

\vspace{-4mm}
\subsection{RSS Model}
\label{Transmission_Channel}
\vspace{-1mm}

As shown in Fig.~\ref{f_system}, the signal received by the user contains a direct line-of-sight (LOS) component and $M$ reflection components. The $m$-th reflection component accounts for the signal transmitted by the AP to the user via the reflection of the $m$-th RIS element. Therefore, the path loss between the AP and the user at the $n$-th block can be expressed as
\vspace{-2mm}
\begin{equation}
  L_{n}(\bm{c})\! =\! s_n(\bm{c})-s^t =\! 20\log_{10}\left|h_{\mathrm{lo}} \!+\!\!\! \sum_{m\in\mathcal{M}}\!\!\! h_{m, n}(c_m)\right|\! + \xi,
  \label{e_received_signal}
  \vspace{-2mm}
\end{equation}
where $s^t$ is the transmission power of AP, $s_n(\bm{c})$ is the RSS at the $n$-th block under configuration $\bm{c}$, $h_{\mathrm{lo}}$ is the channel gain of the direct LOS component, $h_{m, n}(c_m)$ is the gain of the $m$-th reflection channel, and $\xi$ is the log-normal shadowing component which follows Gaussian distribution $\mathcal{N}(0, \sigma^2)$~\cite{ouyang2010received}. 

Based on \cite{goldsmith2005wireless}, $h_{\mathrm{lo}}$ can be expressed as
\vspace{-2mm}
\begin{equation}
  h_{\mathrm{lo}} = \dfrac{\lambda}{4\pi} \cdot \dfrac{\sqrt{g^t_n g^r_n} \cdot e^{- j 2 \pi l_n / \lambda}}{l_n},
  \vspace{-2mm}
\end{equation}
where $\lambda$ is the wavelength of the carrier signal, $g^t_n$ is the power gain of the AP antenna towards the $n$-th block, $g^r_{n}$ is the power gain of the user antenna at the $n$-th block towards the AP, and $l_n$ is distance between the AP and the user at the $n$-th block. Besides, $h_{m, n}(c_m)$ can be expressed as
\vspace{-2mm}
\begin{equation}
  h_{m, n}(c_m) = \dfrac{\lambda}{4\pi} \cdot \dfrac{\sqrt{g^t_m g^r_{m, n}} r_{m}(c_m) e^{- j 2 \pi (l^r_{m}+l^r_{m, n}) / \lambda}}{l^r_{m}l^r_{m, n}},
  \label{equ_mainChannelGain}
  \vspace{-2mm}
\end{equation}
where $g^t_m$ is the power gain of the AP antenna towards the $m$-th RIS element, $g^r_{m, n}$ is the power gain of the user antenna at the $n$-th block towards the $m$-th RIS element, $r_{m, n}(c_m)$ is the reflection coefficient of the $m$-th element in the state $c_m$ for the user at the $n$-th block, $l^r_m$ is the distance between the AP and the $m$-th RIS element, and $l^r_{m, n}$ is the distance between the $m$-th RIS element and the user at the $n$-th block.

Consequently, the probability distribution of RSS value at the $n$-th block under configuration $\bm{c}$ can be expressed as
\vspace{-2mm}
\begin{equation}
  \mathbb{P}(s_n(\bm{c}) = s) = \mathbb{P}(s | \bm{c}, n) = \dfrac{1}{\sqrt{2\pi\sigma^2}}e^{-\dfrac{(s-\mu_n(\bm{c}))^2}{2\sigma^2}},
  \label{e_rss_distribution}
  \vspace{-2mm}
\end{equation}
where $\sigma$ is the standard deviation of the RSS, and $\mu_n(\bm{c})$ is the mean of the RSS, which is defined as $\mu_n(\bm{c}) = s_n(\bm{c}) - \xi$.

\vspace{-3mm}
\section{RIS-aided Multi-user positioning Protocol}
\label{s_protocol}
%\vspace{-1mm}

\begin{figure}[!t]
  \setlength{\abovecaptionskip}{-0pt}
  \setlength{\belowcaptionskip}{-18pt}
  \centering
  \includegraphics[width=2.8in]{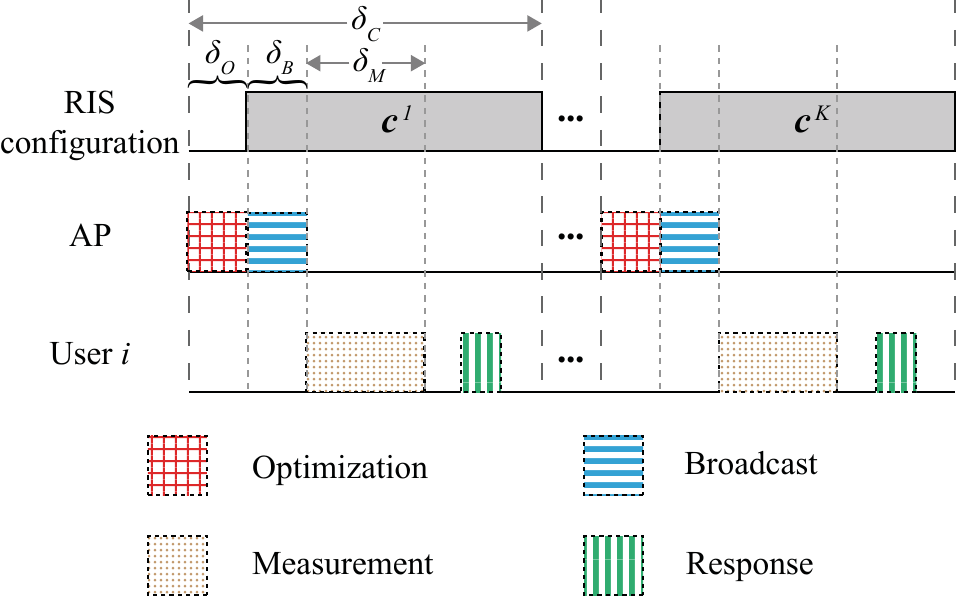}
  \caption{The RIS-aided multi-user positioning protocol.}
  \vspace{-4mm}
  \label{f_sequence}
\end{figure}

In this section, we propose an RIS-aided multi-user positioning protocol, where the configuration is adaptively optimized according to the RSS values of the users. The process of the positioning protocol is illustrated in Fig.~\ref{f_sequence}.

We divide the timeline into cycles with duration $\delta_C$. The positioning process lasts for $K$ cycles, and the positioning results will be broadcast to users when the whole process terminates. Each cycle in the process contains four steps: optimization, broadcast, measurement, and response steps.

\emph{1) Optimization:} In the first $\delta_O$ seconds of the $k$-th cycle, the AP selects the optimal configuration $\bm{c}^k$ for this cycle utilizing the RSS collected in previous cycles. The details of the optimization problem is introduced in the next section.

\emph{2) Broadcast:} In the next $\delta_B$ seconds, the AP broadcasts the configuration $\bm{c}^k$ to all the users and the RIS controller. The RIS controller will change the RIS configuration accordingly.

\emph{3) Measurement:} In this step, the AP sends single-tone signal with frequency $f_c$ for $\delta_M$ seconds, and users record the RSS during this period of time. Let $s^k_i$ denote the average RSS of user $i$ in the $k$-th cycle.

\emph{4) Response:} In the rest of time in this cycle, users need to send the RSS information to the AP. To support multiple users, the time division multiplexing~(TDM) technique is adopted. Specifically, each user is assigned an exclusive time slot and is required to send the signal during the assigned time slot.

\vspace{-3mm}
%%%%%%%%%%%%%%%%%%%%%%%%%%%%%%%%%%%%%%%%%%%%%%%%%%
\section{Problem Formulation}
%%%%%%%%%%%%%%%%%%%%%%%%%%%%%%%%%%%%%%%%%%%%%%%%%%
\label{s_problem}
\vspace{-1.5mm}

%\vspace{-1mm}
\newtheorem{proposition}{\bf Proposition}

In this section, we formulate the optimization problem for the multi-user positioning. To promote the positioning accuracy, we minimize the \emph{positioning loss} in every cycle by selecting a suitable configuration (favorable RSS distribution). To be specific, the positioning loss is the weighted probabilities of false positioning in one cycle, where
\vspace{-2mm}
\begin{equation}
  l(\bm{c}^k) \!=\! \sum_{i \in \mathcal{I}}\sum_{n,n' \in \mathcal{N}, n\ne n'} p^k_{i, n} \gamma^k_{n, n'} \!\int_{\mathcal{R}^k_{i, n'}}\! \mathbb{P}(s^k_i | \bm{c}^k, n) \cdot ds^k_i.
  \label{def_loss}
  \vspace{-2mm}
\end{equation}
Here, $\mathcal{I} = \{1, \cdots, I\}$ denotes the set of users, $p^k_{i, n}$ is the prior probability that user $i$ is at the $n$-th block in the $k$-th cycle, and $\gamma^k_{n, n'}$ is the loss parameter when the positioning result is the $n'$-th block while the user is at the $n$-th block. The integration in (\ref{def_loss}) is the false positioning probability that user $i$ at the $n$-th block is estimated to be at the $n'$-th block. $\mathcal{R}^k_{i, n'}$ is the decision region for the $n'$-th block. That is, if $s^k_i \in \mathcal{R}^k_{i, n'}$, we estimate that user $i$ is at $n'$ in the $k$-th cycle.

The prior probabilities imply our belief about the probability distribution of users' locations based on the RSS values in the previous cycles. According to the Bayes' theorem, the prior probability in the $k$-th cycle can be expressed as
\vspace{-1.5mm}
\begin{equation}
  p^k_{i, n}\!\approx\! \mathbb{P}(n | \bm{c}^{k-1}, s^{k-1}_i)\! =\! \dfrac{p^{k-1}_{i, n} \mathbb{P}(s^{k-1}_i | \bm{c}^{k-1}, n)}{\sum_{n\in\mathcal{N}} p^{k-1}_{i, n} \mathbb{P}(s^{k-1}_i | \bm{c}^{k-1}, n)}.
  \label{d_prior}
  \vspace{-1.5mm}
\end{equation}
Since there is no prior knowledge about the locations of users in the first cycle, we assume that users are uniformly distributed in the SOI, i.e., $p^1_{i, n} = 1/N, \forall n \in \mathcal{N}$.

The loss parameter in the $k$-th cycle is defined as
\vspace{-1.5mm}
\begin{equation}
  \gamma^k_{i, n, n'} = || \bm{r}_n - \bm{r}_{n'} || (1 + \alpha p^k_{i, n'}),\forall n, n'\in\mathcal{N},
  \label{d_loss}
  \vspace{-1.5mm}
\end{equation}
where $\bm{r}_n$ is the location of the $n$-th block’s center, $|| \cdot ||$ denotes the Euclidean distance, and $\alpha$ is a positive parameter. $\gamma^k_{i, n, n'}$ is proportional to $|| \bm{r}_n - \bm{r}_{n'} ||$, which implies that the probability of false positioning decreases when the distance between the estimated block and the correct block increases. Parameter $\gamma^k_{i, n, n'}$ is also positively correlated to $p^k_{i, n'}$, which indicates that if both $p^k_{i, n}$ and $p^k_{i, n'}$ are significant, the false positioning between blocks $n$ and $n'$ has a high weight. Consequently, we need to reduce the probability of false positioning between blocks $n$ and $n'$ by optimizing configurations in order to select the correct block between these two blocks.

The decision region can be obtained using the widely adopted maximum likelihood estimation method~\cite{ouyang2010received}. The decision region for the $n$-th block can be expressed as
\vspace{-1mm}
\begin{equation}
  \mathcal{R}^k_{i, n}\! =\! \{s^k_i | p^k_{i, n}\mathbb{P}(n | \bm{c}^k, s^k_i)\! \ge\! p^k_{i, n'}\mathbb{P}(n' | \bm{c}^k, s^k_i), \forall n'\!\in\!\mathcal{N}/\{n\}\}.
  \label{d_decision}
  \vspace{-1mm}
\end{equation}

Thus, the optimization problem in the $k$-th cycle can be formulated as
\vspace{-2mm}
\begin{subequations}
\begin{align}
  \text{(P1):}\min_{\bm{c}^k}~& l(\bm{c}^k),\\
  s.t.~& (\ref{d_prior})-(\ref{d_decision}),\notag\\
  &c^k_{m} \in \{1, \cdots, C\}, \forall m \in \mathcal{M},\label{p1_c1}
  \vspace{-4mm}
\end{align}
\end{subequations}
where (\ref{p1_c1}) restricts the available states of RIS elements.

\vspace{-3mm}
%%%%%%%%%%%%%%%%%%%%%%%%%%%%%%%%%%%%%%%%%%%%%%%%%%
\section{Algorithm and Analysis}
%%%%%%%%%%%%%%%%%%%%%%%%%%%%%%%%%%%%%%%%%%%%%%%%%%
\label{s_algorithm}
\vspace{-1mm}

In this section, we first propose the CO algorithm to solve the formulated problem, and then analyse the convergence and the complexity of the proposed algorithm. For simplicity, we neglect the superscript $k$ in this section, as the algorithm is the same for each cycle.

Note that the integration in the objective function of (P1) is hard to tackle. To reduce the computational complexity, we use an upper bound provided in the following proposition to replace the integration in the objective function.
\begin{proposition}
  \label{prop_upper}
  An upper bound for the integration in (\ref{def_loss}) can be expressed as
  \vspace{-1.5mm}
  \small\begin{equation}
    \int_{\mathcal{R}_{i, n'}} \mathbb{P}(s_i | \bm{\mu}, n) \cdot ds \le \begin{cases}
      \dfrac{1}{2}e^{-d^2_{i, n, n'}/2},\!\!\!&\!\!\! d_{i, n, n'}\!\ge\!0,\\
      1\!\!-\!\!\dfrac{1}{4}e^{-\dfrac{2d^2_{i, n, n'}}{\pi}},\!\!\!&\!\!\! d_{i, n, n'}\!<\!0,
    \end{cases}
    \vspace{-1.5mm}
    \label{d_upper_integration}
  \end{equation}
  \normalsize
  where \small $d_{i, n, n'} = \dfrac{(\mu_{n'} - \mu_n)^2 - 2\sigma^2 \ln p_{i, n'}/p_{i, n}}{2\sigma|\mu_{n'}-\mu_n|}$.\normalsize
\end{proposition}
\vspace{1mm}
\begin{IEEEproof}
  See Appendix \ref{proof_upper}.
\end{IEEEproof}

Based on (\ref{def_loss}) and (\ref{d_upper_integration}), we have the following remark for the favorable RSS distribution which minimizes the loss in~(\ref{def_loss}).
\newtheorem{remark}{\bf Remark}
\begin{remark}
  In a favorable RSS distribution, 1) if $p_{i, n}, p_{i, n'} \gg 0$, the RSS difference $|\mu_n - \mu_{n'}|$ between blocks $n$ and $n'$ is significant enough to reduce the false
  positioning probability between these two blocks; 2) if $p_{i, n}$ or $p_{i, n'} \approx 0$, the RSS difference $|\mu_{n} -\mu_{n'}|$ can take any value.
\end{remark}

\vspace{-4mm}
\subsection{Configuration Optimization Algorithm}
%\vspace{-1mm}

In this subsection, the CO algorithm based on the global descent method~\cite{ng2007discrete} is proposed to optimize the RIS configuration (RSS distribution) in each cycle.

The CO algorithm contains two phases: initialization and global search phases. In the first phase, a set of configurations is found using the local search method. Since the positioning loss of these configurations may be far away from the global minimum value, in the next phase, the algorithm tries to conduct global search to find configurations with lower values of positioning loss based on the known configurations to improve the algorithm performance. For convenience, two definitions are provided as follows.

\newtheorem{definition}{\bf Definition}

\begin{definition}
  \label{def_un}
  The unit neighborhood of configuration $\bm{c}^*$, $\mathcal{U}(\bm{c}^*)$, is defined by
  \vspace{-1.5mm}
  \begin{equation}
    \mathcal{U}(\bm{c}^*) = \{\bm{c}\ |\ (\bm{c} - \bm{c}^*)\ mod\ C = \pm \bm{e}_m, m \in \mathcal{M}\},
    \vspace{-1.5mm}
  \end{equation}
where \emph{mod} is the modulo operator, and $\bm{e}_m$ is a unit vector. The $m$-th element in $\bm{e}_m$ is $1$, and other elements in $\bm{e}_m$ is $0$.
\end{definition}

\begin{definition}
  \label{def_local}
  The configuration $\bm{c}^*$ is referred to as a local minimum configuration if $l(\bm{c}^*) \le l(\bm{c}) + \epsilon, \forall \bm{c} \in \mathcal{U}(\bm{c}^*)$, where $\epsilon$ is a small but nonzero constant.
\end{definition}

\subsubsection{Initialization Phase}

In this phase, a set of $Z^l$ different local minimum configurations are first obtained using the local minimum configuration search (LMCS) algorithm. The LMCS algorithm can find a local minimum configuration from an initial configuration input using the alternating optimization method. Specifically, if the input configuration $\bm{c}^0$ is not a local minimum configuration, the algorithm will search in $\mathcal{U}(\bm{c}^0)$ to find a configuration $\bm{c}^1$ with the minimum positioning loss. If $\bm{c}^1$ is a local minimum configuration, the algorithm terminates and outputs $\bm{c}^1$. Otherwise, the algorithm will search in $\mathcal{U}(\bm{c}^1)$ to find a new configuration $\bm{c}^2$ and judge whether $\bm{c}^2$ is a local minimum configuration.

The set of $Z^l$ local minimum configurations is denoted by $\mathcal{C}$. These configurations are then sorted in an increasing order according to their positioning loss. 

\subsubsection{Global Search Phase}

In this phase, we iteratively infer other local minimum configurations using $\mathcal{C}$. The method in this phase is inspired by the steepest descent method for the continuous optimization problems. Specifically, three steps are conducted sequentially in each iteration.

\begin{itemize}
  \item The algorithm first computes the descent ratios between the first configuration $\bm{c}^f$ and other configurations in $\mathcal{C}$. The descent ratio is defined as
  \vspace{-1.5mm}
  \begin{equation}
    r(\bm{c}^f, \bm{c}) = \dfrac{l(\bm{c}) - l(\bm{c}^f)}{||\bm{c} - \bm{c}^f||}, \bm{c} \in \mathcal{C}/\{\bm{c}^f\}.
    \vspace{-1.5mm}
  \end{equation}
  The configuration $\bm{c}^m$ with the maximum descent ratio $r(\bm{c}^f, \bm{c}^m)$ is chosen to calculate the steepest descent direction $\bm{d} = (\bm{c}^m - \bm{c}^f)~mod~C$.
  \item Next, the step size $\zeta \in \{1, \cdots, C-1\}$ is enumerated to find a new configuration $((\bm{c}^f + \zeta\bm{d})~mod~C)$ with minimum positioning loss. This configuration will be used as the input of the LMCS algorithm to find a new local minimum configuration $\bm{c}'$.
  \item The configuration $\bm{c}'$ will be inserted into the sorted set $\mathcal{C}$ according to its positioning loss value $l(\bm{c}')$ if $\bm{c}' \notin \mathcal{C}$. Otherwise, a new local minimum configuration $\bm{c}'' \notin \mathcal{C}$ will be found by providing random inputs for the LMCS algorithm, and it will be inserted into the sorted set $\mathcal{C}$.
\end{itemize}
 
The iteration will terminate when $|\mathcal{C}| > Z^u$, and we have $Z^u > Z^l$. After the iteration ends, the first configuration $\bm{c}^f$ in $\mathcal{C}$ is chosen as the output of the CO algorithm. The procedures of the CO algorithm is summarized as Algorithm~\ref{a_co}.

\begin{figure}[!t]
  \removelatexerror
  \begin{algorithm}[H]
    \caption{Configuration Optimization Algorithm}
    \label{a_co}
    \KwIn{Parameter $Z^u$;}
    \KwOut{Configuration $\bm{c}^f$;}
    Initial a set of $Z^l$ different local minimum configurations $\mathcal{C}$ using the LMCS algorithm.\;
    Sort configurations in $\mathcal{C}$ in increasing order of their positioning loss, and $\bm{c}^f$ is the first configuration in $\mathcal{C}$\;
    \While{$|\mathcal{C}| \le Z^u$}{
      Compute the descent ratios $r(\bm{c}^f, \bm{c}), \forall \bm{c} \in \mathcal{C}/\{\bm{c}^f\}$, and find $\bm{c}^m$ with the maximum ratio $r(\bm{c}^f, \bm{c}^m)$\;
      Calculate the steepest descent direction $\bm{d} = (\bm{c}^m - \bm{c}^f)\ mod\ C$\;
      Find $\zeta \in \{1, \cdots, C-1\}$ with minimum loss $l((\bm{c}^f + \zeta\bm{d})~mod~C)$, and calculate the local minimum configuration $\bm{c}'$ using the LMCS algorithm with input $(\bm{c}^f + \zeta\bm{d})$\;
      \eIf{$\bm{c}' \notin \mathcal{C}$}{
        Insert $\bm{c}'$ into sorted set $\mathcal{C}$ according to $l(\bm{c}')$\;
      }{
        Generate a new local minimum configuration $\bm{c}'' \notin \mathcal{C}$, and insert it into $\mathcal{C}$ based on $l(\bm{c}'')$\;
      }
    }
  \end{algorithm}
  \vspace{-4mm}
\end{figure}

\vspace{-5mm}
\subsection{Convergence}
\vspace{-1mm}

\subsubsection{Convergence of the LMCS Algorithm} In each iteration, positioning loss $l(\bm{c})$ is reduced by $\epsilon$ according to Definition~\ref{def_local}. Since $l(\bm{c}) \ge 0$, the algorithm is assured to converge.

\subsubsection{Convergence of the CO Algorithm} In each iteration, a configuration is found using the LMCS algorithm that is guaranteed to converge. Since the CO algorithm contains $(Z^u - Z^l + 1)$ iterations, it is also guaranteed to converge.

\vspace{-5mm}
\subsection{Complexity}
\vspace{-1mm}

\subsubsection{Complexity of the LMCS Algorithm} In the following proposition, an upper bound of positioning loss is provided.
\begin{proposition}
  An upper bound for the positioning loss can be expressed as
  \vspace{-2mm}
  \begin{equation}
    l(\bm{c}) \le l^u = IN(1+\alpha)\sqrt{l^2_x + l^2_y + l^2_z}, \forall \bm{c}.
    \vspace{-1.5mm}
  \end{equation}
\end{proposition}
\begin{IEEEproof}
  The upper bound for the integration (\ref{d_upper_integration}) is not greater than $1$ because $d^2_{i, n, n'} \ge 0$. Besides, the loss parameter $\gamma_{i, n, n'} \le (1+\alpha)\sqrt{l^2_x + l^2_y + l^2_z}$ because probability $p_{i, n'} \le 1$. Based on (\ref{def_loss}), the positioning loss has an upper bound $IN(1+\alpha)\sqrt{l^2_x + l^2_y + l^2_z}$.
\end{IEEEproof}

Since $l(\bm{c}) \ge 0$, the LMCS algorithm has at most $l^u/\epsilon$ iterations. In each iteration, the positioning loss is calculated for $2M$ times. According to (\ref{def_loss}) and (\ref{d_upper_integration}), the complexity of calculating positioning loss is $O(I N^3)$. Therefore, the complexity of the LMCS algorithm is $O(I^2 M N^3)$.

\subsubsection{Complexity of the CO Algorithm} In the CO algorithm, $Z^u$ local minimum configurations are found using the LMCS algorithm, and thus the time complexity is $O(Z^u I^2 M N^3)$.

\vspace{-3mm}
%%%%%%%%%%%%%%%%%%%%%%%%%%%%%%%%%%%%%%%%%%%%%%%%%%
\section{Simulation Results}
%%%%%%%%%%%%%%%%%%%%%%%%%%%%%%%%%%%%%%%%%%%%%%%%%%
\label{s_simulation}
\vspace{-1mm}

In this section, we present the performance of the RIS aided positioning scheme. The layout of the positioning scheme is shown in Fig.~1. The RIS is on the plane $x = 0$, and its center is at $(0, 0, 0)$. The SOI with size $1\times 1 \times 1$m$^3$ is divided into $1,000$ blocks, and its center is at $(1.5, 0, 0)$m. The AP is located at $(0.5, -0.5, 0)$m, and it emits signal with power $s^t = 0$dB and frequency $f_c = 2.4$GHz. The RIS is composed of $64$ elements, and the element separation is $0.06$m. Each element has $4$ states, i.e., $C = 4$, and the amplitudes in different states are set according to the phase shift model in~\cite{abeywickrama2020intelligent}. The antennas equipped by the AP and the users are assumed to be omnidirectional, and we assume their power gains $g^t_n$, $g^r_n$, $g^t_m$ and $g^r_{m, n}$ are equal to $1$. The constants $\alpha = 1,000$, $\epsilon = 0.1$, and the parameters $Z^l = 2$, $Z^u = 5$.

\begin{figure*}[!t]
  \centering
  \subfloat[]{
    %\label{a1} %% label for first subfigure
    \includegraphics[height=1.348in]{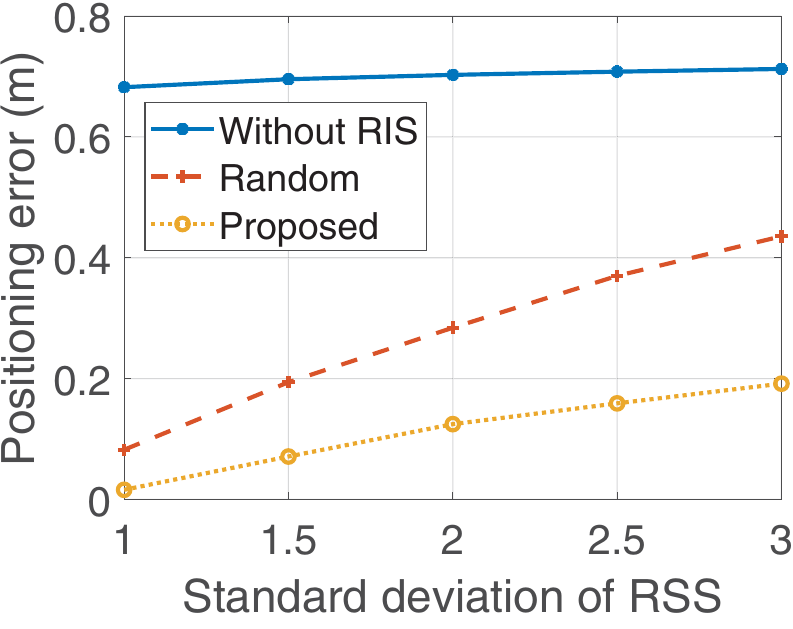}
  }
  \hspace{-0.1in}
  %%----start of second subfigure----
  \subfloat[]{
    %\label{a2} %% label for second subfigure
    \includegraphics[height=1.348in]{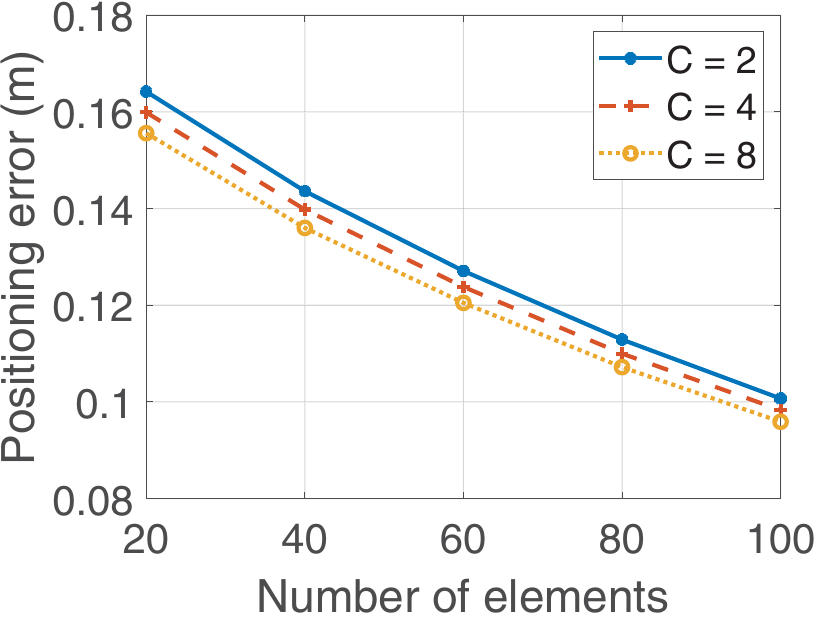}
  }
  \hspace{-0.11in}
    %%----start of second subfigure----
  \subfloat[]{
      %\label{a2} %% label for second subfigure
      \includegraphics[height=1.348in]{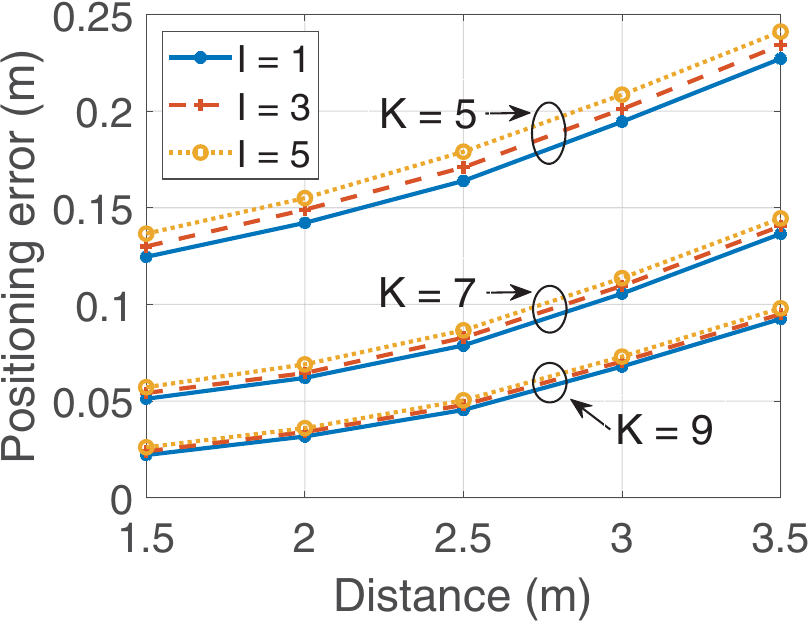}
  }
  \hspace{-0.12in}
    %%----start of second subfigure----
    \subfloat[]{
      %\label{a2} %% label for second subfigure
      \includegraphics[height=1.348in]{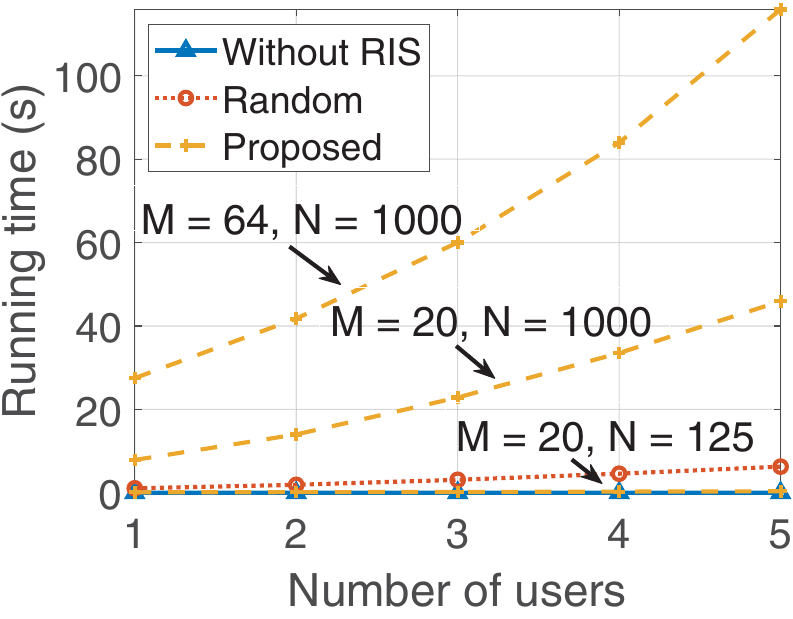}
  }
  \caption{(a) The positioning error $l_e$ versus the standard deviation $\sigma$; (b) The positioning error $l_e$ versus the standard deviation $\sigma$; (c) The positioning error $l_e$ versus the distance between the RIS and the SOI $d_S$; (d) The running time versus the number of users $I$.}
  \label{f_s}
  \vspace{-3mm}
\end{figure*}

To evaluate the positioning accuracy, we define the positioning error $l_e$ as
\vspace{-1.5mm}
\begin{equation}
  l_e = \dfrac{1}{I}\sum_{i\in\mathcal{I}}||\bm{r}^e_i - \bm{r}^g_i||,
  \vspace{-1.5mm}
\end{equation}
where $\bm{r}^e_i$ is the location of the estimated block's center for user $i$, and $\bm{r}^g_i$ is the ground truth. 

For comparison, we also provide the performance obtained by another two schemes: \textbf{1) Without RIS scheme:} In this scheme, the AP equipped with multiple antennas are utilized to create different signal patterns, and users' locations can be obtained by comparing the RSS with the signal patterns~\cite{ohtani2014evaluation}. \textbf{2) Random configuration scheme:} In this scheme, the configurations are set randomly in different cycles.

Fig.~\ref{f_s}(a) illustrates the positioning error $l_e$ versus the standard deviation of RSS $\sigma$ when $I = 1$. In the without RIS scheme, the number of antennas equipped by the AP is 2, and we set the number of RIS elements $M = 64$ in the random configuration and the proposed schemes. We can observe that $l_e$ obtained by the proposed scheme is much lower than that obtained by the other three schemes, which verifies the effectiveness of the proposed scheme. Besides, we can observe that $l_e$ increases with $\sigma$ for all the schemes. Since the standard deviation is negatively related to the measurement time $\delta_M$~[3], the positioning performance can be improved by extending the measurement time.

Fig.~\ref{f_s}(b) depicts the positioning error $l_e$ versus the number of elements $M$ when $I = 1$.\footnote{We use $1$, $2$, and $3$ bit coded RISs, and thus the number of states $C = 2, 4, 8$.} We can observe that the error $l_e$ decreases when the number of elements $M$ and the number of states $C$ increases. According to (2) and (4), by increasing $M$ or $C$, the path loss can be adjusted in a larger range with finer resolution, and thus, the RSS distributions with lower positioning loss can be created by the RIS.

Fig.~\ref{f_s}(c) depicts the positioning error $l_e$ versus the distance between the RIS and the SOI $d_S$. We can observe that $l_e$ increases when $d_S$ increases or $K$ decreases. Besides, $l_e$ also increases with the number of users $I$. This is because the average RSS variance for each user under different configurations declines when more users need to be considered simultaneously in the optimization problem.

Fig.~\ref{f_s}(d) depicts the running time versus the number of users $I$. We can observe that the running time of all the schemes increases with the number of users, and the proposed scheme has the highest complexity. However, we can also observe that the running time of the proposed scheme decreases when the number of elements and blocks decreases, which implies that we can limit the number of elements or the size of the SOI to ensure acceptable complexity.

\vspace{-2mm}
%%%%%%%%%%%%%%%%%%%%%%%%%%%%%%%%%%%%%%%%%%%%%%%%%%
\section{Conclusion}
%%%%%%%%%%%%%%%%%%%%%%%%%%%%%%%%%%%%%%%%%%%%%%%%%%
\label{s_conclusion}
\vspace{-1mm}

In this letter, we have studied the RIS-aided multi-user wireless indoor positioning using the RSS based technique. We have proposed an RIS-aided multi-user positioning protocol and formulated the optimization problem for the multi-user positioning. The CO algorithm has been designed to solve the formulated problem, and its effectiveness has been verified by the simulation results. It can also be concluded from the simulation results that the positioning error increases when the standard deviation of the RSS, the distance between the RIS and the SOI, and the number of users increase, or the number of RIS elements, the number of element states, and the number of cycles decreases.

\vspace{-3mm}
\appendices
\section{Proof of Proposition~\ref{prop_upper}}
\label{proof_upper}
\vspace{-2mm}

Based on (\ref{e_rss_distribution}), decision region $\mathcal{R}_{i, n'}$ can be expressed as
\vspace{-1.5mm}
\begin{equation}\small
  \mathcal{R}_{i, n'}\!\! =\!\! \left\{\!s_i\!:\!\! (s_i \!-\!\! \mu_{n'})^2 \!\!-\! (s_i \!-\!\! \mu_{n})^2 \!\!\le\!\! 2\sigma^2 \!\ln\! \dfrac{p_{i, n'}}{p_{i, n}},\! n\!\in\!\mathcal{N}\!/\!\{n'\}\!\right\}\!.
  \vspace{-1.5mm}
\end{equation}

The union bound method~\cite{goldsmith2005wireless} can provide a tight upper bound for region $\mathcal{R}_{i, n'}$ with a high SNR. The region $\mathcal{R}_{i, n', n}$ provided by the union bound method can be expressed as
\vspace{-3mm}
\small\begin{align}
  \mathcal{R}_{i, n'\!, n}\! =&\! \left\{\!s_i\!:\! (s_i - \mu_{n'})^2 - (s_i - \mu_{n})^2 \le 2\sigma^2 \ln p_{i, n'}/p_{i, n}\right\}\notag\\
  =&\! \left\{\! s_i\!:\! (s_i \!-\! \mu_n)(\mu_{n'}\!-\!\mu_n) \!\ge\! \dfrac{1}{2}(\mu_{n'} \!-\! \mu_n)^2 \!-\! \sigma^2\! \ln \dfrac{p_{i, n'}}{p_{i, n}}\!\right\}
  \vspace{-3mm}
\end{align}
\normalsize

According to (\ref{e_rss_distribution}), $(s_i - \mu_n)(\mu_{n'}-\mu_n)$ follows the Gaussian distribution $\mathcal{N}(0, (\mu_{n'}-\mu_n)^2\sigma^2)$. Thus, the integration in (\ref{def_loss}) can be expressed as
\vspace{-3mm}
\small\begin{align}
  \int_{s_i \in \mathcal{R}_{i, n', n}} \!\!\!\!\!\!\!\mathbb{P}(s_i | \bm{c}, n) ds \!=\! &~Q(d_{i, n, n'})
  \!\le\!
  \begin{cases}
    \dfrac{1}{2}e^{-d^2_{i, n, n'}/2},\!\!\!&\!\! d_{i, n, n'}\!\ge\!0,\\
    1\!\!-\!\!\dfrac{1}{4}e^{-\dfrac{2d^2_{i, n, n'}}{\pi}},\!\!\!&\!\! d_{i, n, n'}\!<\!0,
  \end{cases}
  \vspace{-3mm}
\end{align}
\normalsize
where $d_{i, n, n'} = \dfrac{(\mu_{n'} - \mu_n)^2 - 2\sigma^2 \ln p_{i, n'}/p_{i, n}}{2\sigma|\mu_{n'}-\mu_n|}$, and the relationship provided in~\cite{goldsmith2005wireless} is utilized in the last step.

\end{document}